\documentclass[doublespacing]{elsart}

\usepackage{longtable}
\usepackage{rotating}
\usepackage[T1]{fontenc}

\usepackage{graphicx}
\usepackage{amssymb}
\providecommand{\tabularnewline}{\\}

\begin{document}

\begin{frontmatter}


 \title{Optical Properties of Bialkali Photocathodes}
 \author{D. Motta\corauthref{cor1}}
 \ead{dario.motta@mpi-hd.mpg.de}
 \corauth[cor1]{Corresponding author}
 \author{and S. Sch\"onert}
 \address{Max-Planck-Institut für Kernphysik, Saupfercheckweg 1, 69117 Heidelberg, Germany}


\begin{abstract}
The optical properties of the ``bialkali'' \emph{KCsSb} and \emph{RbCsSb}
photomultiplier cathodes have been experimentally investigated in the visible range.
The measurements carried out include the absolute reflectance at near-normal
incidence, the polarization-dependent relative reflectance at various angles
and the change in polarization upon reflection from the photocathode. These experimental 
inputs have been combined with a theoretical model to determine the complex refractive
index of the photocathodes in the wavelength range $[380\, nm,\,680\, nm]$
and their thickness. As a result of this work, we derive a model which predicts
the fraction of light impinging on a photomultiplier tube that
is reflected, absorbed or transmitted, as a function of wavelength
and angle, and dependent on the medium to which the photomultiplier is coupled.
\end{abstract}

\begin{keyword}
photomultiplier tubes \sep bialkali photocathodes \sep optical properties

\PACS 85.60.Ha  \sep 78.20.Ci \sep 78.66.-w

\end{keyword}

\end{frontmatter}

\section{\label{sec:Introduction}Introduction}

\begin{figure} 
\begin{center}
 \includegraphics[width=1.0\columnwidth,
  keepaspectratio]{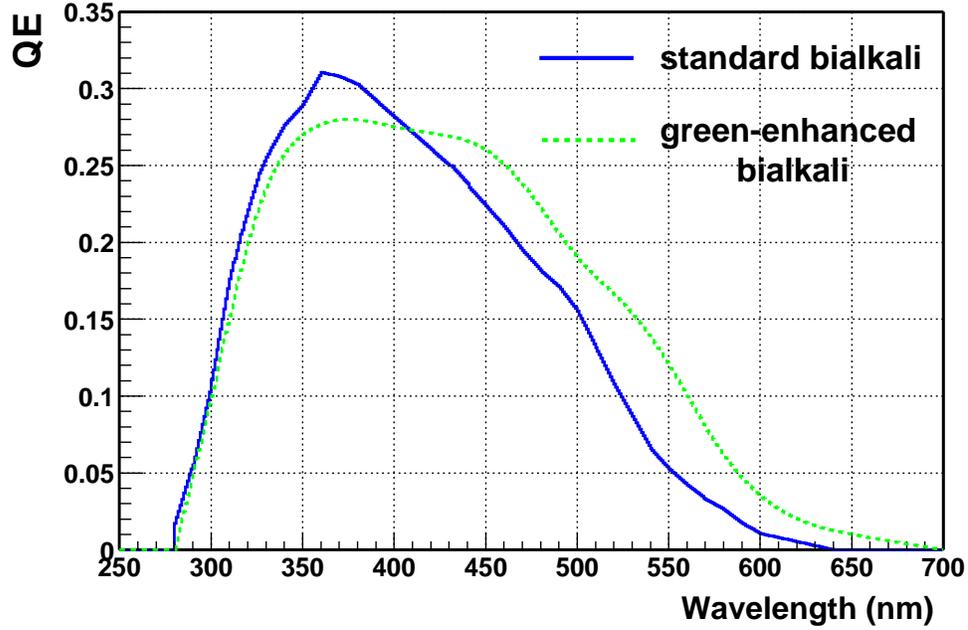}
\end{center}

\caption{\textsl{\small \label{fig:QE-spectra}Typical spectral response
of blue-sensitive and green-enhanced bialkali photocathodes. The ordinate
gives the probability for a photon to produce a photoelectron. The
curves are usually measured by illuminating the PMT at the center
in air, at normal incidence. The cut-off at $\sim310\, nm$ is due
to the absorption by the PMT glass window.}}
\end{figure}

Photomultiplier tubes (PMTs) with bialkali photocathodes are widely
used in astrophysics, nuclear and particle physics. In many cases,
the only optical property of a PMT of interest for the design of an experiment is its spectral
sensitivity, \emph{i.e.} the fraction of photons converted into detected
photoelectrons as a function of the wavelength (Fig. \ref{fig:QE-spectra}).
Recent developments in particle physics, especially in neutrino physics,
have led to the construction or proposal of several multi-ton up to
megaton detectors, in which the interactions are detected either via
Cherenkov light in water (or ice) or via scintillation light from
organic liquid scintillators \cite{Borexino-2002,Double-Chooz,KamLAND-results,SK_detector,SNO,Nu2004}.
Data analysis in these experiments is based on the comparison with the
predictions of Monte Carlo simulations, which typically do not include 
a detailed description of light interaction with the PMTs. 

Three processes are possible for a photon impinging on a PMT:

\begin{enumerate}
\item Absorption in the photocathode, with probability $A(\lambda,\theta)$
\item Reflection from the PMT window or photocathode, with probability
$R(\lambda,\theta)$
\item Transmission inside the PMT, with probability $T(\lambda,\theta)=1-A(\lambda,\theta)-R(\lambda,\theta)$
\end{enumerate}
In case of absorption, a photoelectron is produced, which has a certain
probability to escape the layer towards the interior of the PMT, to be
accelerated to the first dynode and start an avalanche, resulting
in a detectable signal. It is convenient to express the probability
that a photon contributes to a signal as the product of two probabilities:\begin{equation}
QE(\lambda,\theta)=A(\lambda,\theta)\times P_{conv}(\lambda)\label{eq:QE}\end{equation}
where $A(\lambda,\theta)$ is the probability that the photon is absorbed
in the photocathode, which is a function of the wavelength $\lambda$
and the incidence angle $\theta$, and $P_{conv}$ is the conversion
factor for such absorption to result in an avalanche. $P_{conv}$
depends on $\lambda$ (due to the different kinetic energy transferred
to the photoelectron in the absorption process), and - not included in
Eq. \ref{eq:QE} - on the location of the photon absorption on
the PMT surface and the operating high voltage. $P_{conv}$ is here assumed
independent of $\theta$, as demonstrated in \cite{Lay}.

The probability $QE(\lambda,\theta)$ is the \emph{Quantum Efficiency}. 
In Fig. \ref{fig:QE-spectra} the typical QE spectra are shown for two of the 
most commonly used photocathodes: the \emph{blue-sensitive bialkali} (\emph{KCsSb}) 
and the \emph{green-enhanced bialkali} (\emph{RbCsSb}). 
The curves are usually measured in air
for light impinging at near normal incidence ($\theta\sim0)$. 

For most applications the information in Fig. \ref{fig:QE-spectra}
is sufficient, since it characterizes the PMT detection efficiency
for ``standard'' operations. However, the physics of light interaction
with the PMT is more complex. First, the QE is not only a function
of $\lambda$, but also of $\theta$; second, in many detectors the
PMTs are typically coupled to media with $n>1$, and not to air; last,
the probabilities $R(\lambda,\theta)$ and $T(\lambda,\theta)$ are
not implicit in Eq. \ref{eq:QE}. 

A precise knowledge of the photocathode optical properties would allow 
the functions $QE(\lambda,\theta)$, $R(\lambda,\theta)$  and 
$T(\lambda,\theta)$ to be predicted for a PMT in contact with any medium. 
In spite of the great importance of this issue for detector-modeling, 
surprisingly very little is documented in the literature. 
A major breakthrough was accomplished by Moorhead
and Tanner \cite{Moorhead} (hereafter referred to as M\&T), who measured
the optical constants of a \emph{KCsSb} bialkali photocathode in a EMI 9124B PMT at the
wavelength of $\lambda=442\, nm$, for implementation in the MC simulation
of the SNO experiment \cite{SNO}. The method they used gave them the ability
to break the ambiguity in the optical constants typical of this kind of 
determinations. In their work M\&T also tried to deduce the wavelength 
dependence of the photocathode refractive index, by reinterpreting and 
reanalyzing earlier measurements at few other wavelengths in the visible 
spectrum reported in \cite{Timan}. Later on, Lang used M\&T's technique
to check their measurements on the same photocathode and to extend
the investigation to several other photocathodes \cite{Lang}, including
the \emph{RbCsSb}. Lay subsequently exploited the same technique to study the
performance of several Hamamatsu R1408 PMTs used in SNO \cite{Lay}.
However, all of the above measurements were limited to the single
wavelength $\lambda=442\, nm$.

The objective of this work was to extend the existing measurements
on the \emph{KCsSb} and \emph{RbCsSb} photocathodes by covering the
whole visible spectrum and consequently to predict the interaction of visible
light with bialkali PMTs in any experimental condition. The two investigated
photocathodes are the most usual choice in particle physics. The former
is suitable for the detection of Cherenkov light and scintillation
light from primary fluors, the latter ensures good blue-green sensitivity
and matches better the emission spectra of secondary fluors (wavelength-shifters)
in three-component organic scintillator mixtures. 

The original motivation for this work was to study the impact on the solar 
neutrino Yb-LENS experiment \cite{Raju97,Stefan-Osaka,LENS-nu02} of 
light reflections from PMTs. The signature of a $\nu_{e}$ interaction
with the target nuclide $^{176}Yb$ is a delayed $e^{-}-\gamma$ coincidence with 
life time $\tau=50\, ns$. A modular design is envisaged, with Yb dissolved in an organic
liquid scintillator. In the investigated detector concept, a fraction of the
primary scintillation light reflects off the PMTs and is eventually detected at
a later time, with a delay given by the time of propagation through a module.
This delay is of the order of $\sim30\, ns$ for a typical module length of $\sim4\, m$,
hence PMTs-reflected light falls in the time-window of the $\nu$-signature.
Consequently, reflections contribute to increase
the probability that single events are misidentified as correlated
double events, \emph{i.e}. $\nu_{e}$-candidates \cite{mythesis}. 

More in general, the study of light reflections from the PMTs is of concern 
in all detectors where this reflected light can cause a delayed signal.
For example, in Borexino (and similarly in KamLAND) reflections
weaken the $\alpha-\beta$ discrimination power, which is based on
the fact that the excitation of organic scintillators from $\alpha$-particles 
create a larger fraction of late light compared to the excitation from electrons \cite{Borexino-2002}.

Another application of the optical model of PMTs is the prediction of
the angular dependence of the QE, for a PMT operated in air as well as in liquid.

\section{Theory and Methodology}

\subsection{Introduction}

\label{sec:Theory-and-methodology}The photocathode of a PMT is
a thin layer of a multi-alkali semiconducting alloy, which is evaporated
onto the back side of the glass window during production. Light impinging
on a photocathode is in part reflected, in part absorbed and in part
transmitted. The photocathode thickness is a compromise aiming at
maximizing the probability that a photon results in a signal: if the
layer is too thin, little light can be absorbed; if it is too thick,
the resulting photoelectrons cannot efficiently escape from the photocathode.

The description of such systems is covered by the optics
of thin absorbing films, a wide field on which a very rich literature
exists (see \emph{e.g.} the textbook \cite{Heavens}). The theoretical model
merges the optical description of an absorbing material, through its
complex refractive index, and the optics of thin layers, where the 
wave fronts of different orders of reflection are added
coherently to determine the resulting reflected wave amplitude. Assuming
that the surface of a photocathode is sufficiently regular and uniform that
the model of thin films is applicable, then the functions $A(\lambda,\theta)$,
$R(\lambda,\theta)$, $T(\lambda,\theta)$ are predictable, once the three parameters of the problem are known: real
and imaginary part of the refractive index, $n^{*}=n+ik$ (\emph{i}
imaginary unit, both \emph{n} and \emph{k} function of the wavelength)
and thickness \emph{d} of the layer. The challenge is to invert the
results of optical measurements performed under certain specific conditions
(angle of incidence, polarization, refractive index of the medium
coupled to the PMT) in order to infer the constants \emph{n}, \emph{k}
and \emph{d}. This can be done by fitting the experimental results
with the optical model of the photocathode, with \emph{n}, \emph{k}
and \emph{d} as free parameters.

Unfortunately the problem has often a high degeneracy: for a single
measurement (for example absolute reflectance at a certain incidence
angle) there exist infinite solutions for (\emph{n,k,d}) that reproduce
the data. For each considered wavelength at least three independent measurements
are needed, but this is often still insufficient to guarantee the
convergence to a unique solution.

Several techniques have been developed for the study of thin solid
films (see \cite{Heavens,Tompkins} for a review). However,
the photocathode of a PMT is a challenging sample. It cannot be directly
studied, because it is not chemically stable when exposed to air.
Therefore some optical observables, like reflectance from
the photocathode side and its transmittance, cannot be measured. Furthermore,
polarization analysis of the light reflected off the photocathode
can be disturbed by the birefringence that is induced on the glass
window by the mechanical stress due to the pressure difference (the
interior of the PMT is under vacuum). The limitations due to these 
experimental difficulties are discussed in Secs. \ref{sec:Opt-meas-photocath}
and \ref{sec:Results}.

\subsection{The Model}

Dealing with light reflection from a PMT, four regions can be defined,
each characterized by its own refractive index: the medium where the
light originates, with $n_{1}$; the PMT glass envelope, with $n_{2}$;
the photocathode, with $n_{3}^{*}=n_{ph}+ik_{ph}$, and the vacuum
inside the PMT, with $n_{4}=1$. The amplitudes of the reflected and
transmitted waves at the interface glass-photocathode ($n_{2}$ to
$n_{3}^{*}$) are given by the following formulas (see \emph{e.g.} \cite{Optics}
for a derivation):\begin{equation}
\begin{array}{l}
a{}_{R}(\lambda,\theta)=r_{23}+\frac{t_{23}t_{32}r_{34}\exp(-2i\delta)}{1+r_{23}r_{34}\exp(-2i\delta)}\\
\\a{}_{T}(\lambda,\theta)=\frac{t_{23}t_{34}\exp(-i\delta)}{1+r_{23}r_{34}\exp(-2i\delta)}\end{array}\label{eq:R&T_amplitudes_thin_film}\end{equation}
where\begin{equation}
\begin{array}{l}
r_{ij}=\frac{n_{i}\cos\left(\theta_{i(j)}\right)-n_{j}cos\left(\theta_{j(i)}\right)}{n_{i}\cos\left(\theta_{i(j)}\right)+n_{j}cos\left(\theta_{j(i)}\right)}\\
\\t_{ij}=\frac{2n_{i}\cos\left(\theta_{i}\right)}{n_{i}\cos\left(\theta_{i(j)}\right)+n_{j}cos\left(\theta_{j(i)}\right)}\\
\\\delta=\frac{2\pi dn_{3}^{*}}{\lambda}\cos\left(\theta_{3}\right)\end{array}\label{eq:R&T_aux}\end{equation}
In Eq. \ref{eq:R&T_aux}, $n_{l}$ is the refractive index of
the $l^{th}$ region, $\theta_{k}$ the angle of the propagating light
beam with respect to the normal in the same region. This is calculated using
Snell's law starting from the angle of incidence on the PMT window,
$\theta_{1}\equiv\theta$. All functions and variables in Eq. \ref{eq:R&T_amplitudes_thin_film}
and \ref{eq:R&T_aux} are complex and the imaginary part of $n_{1}$, $n_{2}$
and $n_{4}$ is set to 0. Eq. \ref{eq:R&T_amplitudes_thin_film}
holds for both light polarizations, perpendicular and parallel
relative to the photocathode plane (defined as \emph{s} and \emph{p} waves, respectively),
provided that the definitions for $r_{ij}$ and $t_{ij}$ are changed
by swapping the \emph{i} and \emph{j} indices as indicated in Eq.
\ref{eq:R&T_aux} (the formulas with first indices apply to the
p-wave). 

Eq. \ref{eq:R&T_amplitudes_thin_film} can be used to predict
the total reflectance of the PMT by adding the Fresnel reflection
at the medium-to-window interface ($n_{1}$ to $n_{2}$) to the reflectance
from the photocathode:\begin{equation}
\begin{array}{l}
R_{tot}(\lambda,\theta)=\frac{1}{2}\left[R_{s}^{tot}(\lambda,\theta)+R_{p}^{tot}(\lambda,\theta)\right]\\
\\R_{s,p}^{tot}(\lambda,\theta)=F_{s,p}+\frac{R_{s,p}\left(1-F_{s,p}\right)^{2}}{1-F_{s,p}R_{s,p}}\\
\\R_{s,p}(\lambda,\theta)=\left|a{}_{R}^{s,p}\right|^{2}\end{array}\label{eq:Refl_thin_film}\end{equation}
where the quantities:\begin{equation}
\begin{array}{l}
F_{s}(\lambda,\theta)=\left[\frac{\sin\left(\theta_{1}-\theta_{2}\right)}{\sin\left(\theta_{1}+\theta_{2}\right)}\right]^{2}\\
\\F_{p}(\lambda,\theta)=\left[\frac{\tan\left(\theta_{1}-\theta_{2}\right)}{\tan\left(\theta_{1}+\theta_{2}\right)}\right]^{2}\end{array}\label{eq:Fresnel_s,p}\end{equation}
represent the Fresnel reflection coefficients at the interface with
the PMT window (see \cite{Optics}). Similarly for the transmittance:\begin{equation}
\begin{array}{l}
T_{tot}(\lambda,\theta)=\frac{1}{2}\left[T_{s}^{tot}(\lambda,\theta)+T_{p}^{tot}(\lambda,\theta)\right]\\
\\T_{s,p}^{tot}(\lambda,\theta)=\frac{T_{s,p}\left(1-F_{s,p}\right)}{1-F_{s,p}R_{sp}}\\
\\T_{s,p}(\lambda,\theta)=\frac{n_{4}\cos\left(\theta_{4}\right)}{n_{2}\cos\left(\theta_{2}\right)}\left|a{}_{T}^{s,p}\right|^{2}\end{array}\label{eq:Tran_thin_film}\end{equation}
The last optical function, absorption, is deduced from Eqs. \ref{eq:Refl_thin_film}
and \ref{eq:Tran_thin_film}, by using the identity:\begin{equation}
A(\lambda,\theta)=1-R(\lambda,\theta)-T(\lambda,\theta)\label{eq:Abs_thin_film}\end{equation}

\subsection{\label{sub:Experimental-Approaches}Experimental Approaches}

\subsubsection*{Preliminary Remarks}

As a consequence of Snell's law, the angle at which a ray of light with
a given incidence angle emerges
after crossing a series of parallel layers depends only on the refractive
index of the first and last medium. If $n_{1}\simeq1$ (PMT in air) there is no critical angle for
total internal reflection. The part of the PMT reflectance due to
the direct Fresnel contribution of the window is trivial. In a large
region of the (\emph{n,k,d}) parameter space the photocathode reflectance\footnote
{Here and throughout this paper we call \emph{photocathode reflectance} 
the fraction of the light incident on the PMT that is reflected by the photocathode alone; 
\emph{PMT reflectance} the fraction of the light that is reflected by the PMT as a whole,
including the photocathode reflectance and the Fresnel reflection at the window.}
for both polarizations (given by Eqs. \ref{eq:Refl_thin_film}
after subtraction of the window contribution $F_{s,p}$) is nearly
constant for $\theta$$\lesssim60^{\circ}$ and then drops to zero
(simply because the Fresnel reflection at the window rapidly increases
to $100\,\% $ for $\theta$ approaching $90^{\circ}$, so that less light
can refract and reach the photocathode). Therefore, reflectance measurements
in air, even at different angles, provide a weak constraint on the
solution of the fit, as pointed out by M\&T in \cite{Moorhead}. 

On the other hand, if $n_{1}>1$ (for example PMT in water or scintillator),
then there is a critical angle $\theta_{c}$ above which no light
can be transmitted. In this case the photocathode reflectance shows
a plateau until $\theta_{c}$ and then a peculiar shape, very different
for parallel and perpendicular polarizations. The optical model of
the photocathode shows that, in this case, the free parameters (\emph{n,k,d})
are strongly constrained by the angular dependence of the reflectance,
in particular by the shape around $\theta_{c}$. 

\begin{figure} 
\begin{center}
 \includegraphics[width=1.0\columnwidth,
  keepaspectratio]{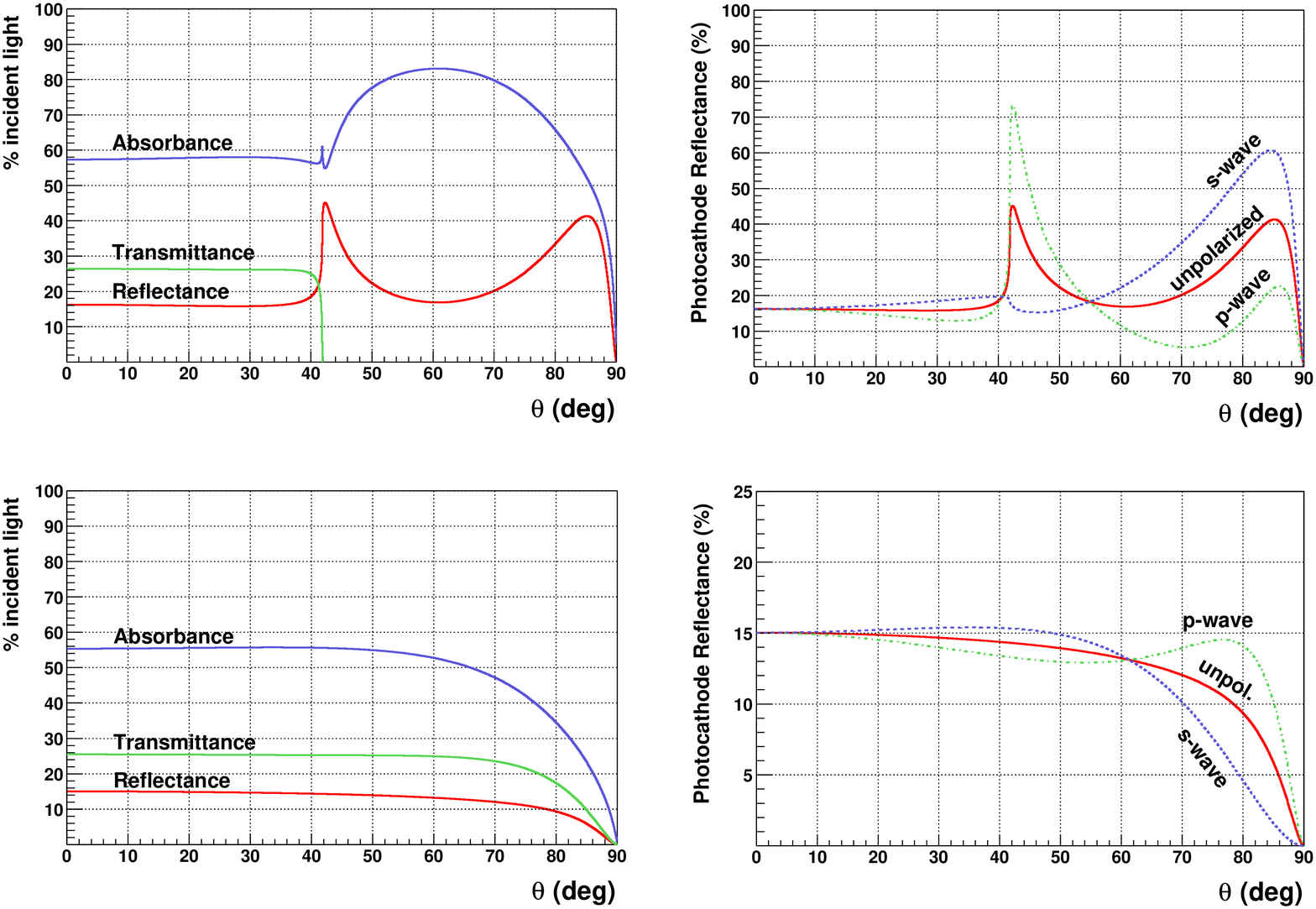}
\end{center}

\caption{\textsl{\small \label{fig:ART}}}

\textbf{\textsl{\small Top left}}\textsl{\small : predictions for the photocathode 
absorption, reflectance and transmittance, assuming the following parametrization:
$n_{1}=1.5$ (scintillator), $n_{2}=1.51$ (glass), $n_{3}=2.7+1.5i$
(photocathode), $n_{4}=1$ (vacuum); $d=20\, nm$; $\lambda=442\, nm$. }{\small \par}

\textbf{\textsl{\small Top right}}\textsl{\small : reflectance for
unpolarized light, s and p waves for the same case.}{\small \par}

\textbf{\textsl{\small Bottom left}}\textsl{\small : predictions for
the same parameters, but PMT in air: $n_{1}=1$. }{\small \par}

\textbf{\textsl{\small Bottom right}}\textsl{\small :} \textsl{\small reflectance
for unpolarized light, s and p waves for the same case.}{\small \par}

\textsl{\small In all the plots the ordinates are given in percent
of the incident light, and the first order Fresnel reflection from
the glass is not included (for this reason the sum $A+R+T$ drops
to zero, for $\theta\rightarrow90^{\circ}$). }
\end{figure}

The different behavior of a PMT in air and in a $n\sim1.5$ medium is 
illustrated with an example in Fig. \ref{fig:ART}.

\subsubsection*{Previous Measurements}

Based on the above observations, M\&T have built an experimental set-up to measure
the photocathode reflectance as a function of the angle of incidence
for a PMT immersed in water, at the wavelength of $\lambda=442\, nm$,
provided by a \emph{He-Cd} laser. With this technique they were able
to establish unambiguously the three optical constants of the \emph{KCsSb}
bialkali photocathode they investigated: $n=2.7\pm0.1$, $k=1.5\pm0.1$,
$d=(23\pm2)\, nm$ \cite{Moorhead}. The result was obtained by fitting
the data with the same optical model summarized in Eqs. \ref{eq:R&T_amplitudes_thin_film}
to \ref{eq:Fresnel_s,p}. The goodness of the best fit was sufficient
to demonstrate that the assumed theoretical model is an acceptable
description of the photocathode. Later on Lang and Lay measured
several PMTs equipped with the \emph{KCsSb} photocathode at the same 
wavelength and found that the above parameters are quite typical,
though PMT-to-PMT deviations may be larger than the statistical error
\cite{Lay,Lang}.

\subsubsection*{This Work}

Our aim was to perform a spectroscopic investigation of the optical
properties of bialkali photocathodes, covering the whole visible spectrum. For this
purpose we used optical intrumentation designed to operate in air and our strategy was
to break the degeneracy discussed earlier by making a global
fit to a set of independent measurements. This kind of approach is
quite ``standard'' in the study of thin solid films, as discussed
in \cite{Tompkins}. The optical constants measured at $\lambda=442\, nm$
by M\&T, Lang and Lay give a further contraint to break the residual ambiguities.
For a photocathode of a PMT, light reflected from the glass window
side is the only accessible observable. In addition to reflectance
determinations, we also performed \emph{ellipsometric} measurements. 

Ellipsometry (see \cite{Tompkins}) is
based on the measurement of the change in the polarization state of
light upon reflection off a plane surface. An ellipsometer
consists of a polarizer and an analyzer: light is linearly polarized,
sent to a sample and finally the reflected beam is analyzed. In general
the reflection changes the polarization from linear to elliptic, in
a way that depends on the three optical constants (\emph{n,k,d}).
Any linear polarization can be thought of as the superposition in
phase of two orthogonal components. The frame of reference is chosen
relative to the sample plane, which defines the parallel \emph{p}-\emph{wave},
and the perpendicular \emph{s-wave}. The reflection off a surface
changes the relative intensity ratio between \emph{p} and \emph{s}
waves and introduces a phase delay. This change can be expressed by the ratio
of the complex reflection amplitudes, which translates to an angle
and a phase:\begin{equation}
\frac{a{}_{R}^{p}}{a{}_{R}^{s}}=\tan\Psi e^{i\Delta}\label{eq:ellipsometry}\end{equation}
 where the amplitudes $a{}_{R}^{p}$ and $a{}_{R}^{s}$ are given by
the first row of Eq. \ref{eq:R&T_amplitudes_thin_film}. The angle 
$\Psi$ and the phase $\Delta$ are the two parameters measured by ellipsometry. 

Ellipsometry is one of the most sensitive methods for the analysis
of thin films and has the advantage over conventional reflectometry
to be independent of absolute determinations, where systematics are
difficult to control. To our knowledge, this technique has been applied
here for the first time to the study of PMT photocathodes.

\section{\label{sec:Opt-meas-photocath}Optical Measurements of Photocathodes}

Two $1.5^{\prime\prime}$ PMTs from ETL (Electron Tubes Limited) have
been investigated, 9102B and 9902B, the former equipped with a blue-sensitive
\emph{KCsSb} bialkali photocathode, the second with a green-enhanced 
\emph{RbCsSb} bialkali photocathode. Both PMTs have a plano-plano
window. The measurements performed are:

\begin{enumerate}
\item Absolute reflectance for unpolarized light at $\theta=7^{\circ}$
in the wavelength range $[250\, nm-700nm]$
\item Relative reflectance for \emph{p} and \emph{s} polarizations at $\theta=45^{\circ},\:55^{\circ},\:65^{\circ}$
in the wavelength range $[400\, nm-700nm]$
\item Ellipsometric measurement of $(\Psi,\Delta)$ in the wavelength range
$[420\, nm-740nm]$
\end{enumerate}
For the cases 2 and 3, the lower limit of the probed wavelength range
is the shortest allowed by the instrument or the adopted measurement
procedure, as will be explained in the next sections. The upper limit 
is arbitrarily chosen, well above the cut-off of the PMT sensitivity 
(see Fig. \ref{fig:QE-spectra}).

\subsection{Absolute Reflectance at a Fixed Angle}

\subsubsection*{Experimental Technique}

\begin{figure} 
\begin{center}
 \includegraphics[width=1.0\columnwidth,
   keepaspectratio]{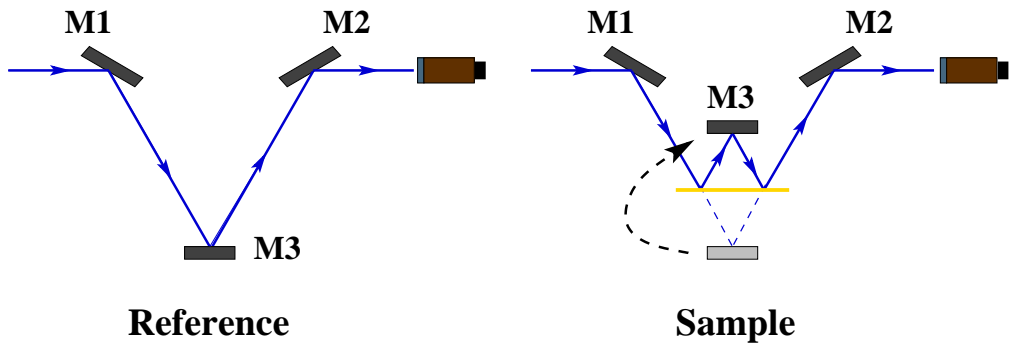}
\end{center}

\caption{\textsl{\small \label{fig:V-W}Schematic diagram of the V-W technique
applied to measure the absolute reflectance of our sample PMTs. The
light intensity with the apparatus in the ``Sample'' configuration
($I_{S}$) is compared to the one measured in the ``Reference''
configuration ($I_{R}$). The mirrors M1 and M2 are in a fixed position,
while the mirror M3 can be moved as shown on the right side of the
figure. The sample to measure is located exactly half way between
the two slots for M3. The optical length is the same in the two configurations,
as well as the reflectance at the mirrors M1, M2 and M3. The only
difference is that light in the right side configuration must reflect
twice on the sample. Thus: $R=\sqrt{I_{S}/I_{R}}$. The result is
an average of the sample reflectance in the two illuminated spots.
In the V-W accessory of the Varian Cary 400 spectrophotometer the
angle of incidence on the sample is $7^{\circ}$ and the instrument
is designed to operate in air.}}
\end{figure}

The PMT absolute reflectance has been measured in air, at near normal incidence 
($\theta=7^{\circ}$), with a V-W accessory of a UV/Visible Varian Cary 400 
spectrophotometer. The V-W technique (Fig. \ref{fig:V-W} and relevant caption) 
allows the measurement of the absolute specular reflectance of a flat sample
without the use of any reference mirror. A precision of $\sim\pm1\,\%$
is attainable (instrument specification).

\subsubsection*{Results and Discussion}

\begin{figure} 
\begin{center}\
 \includegraphics[width=1.0\columnwidth,
  keepaspectratio]{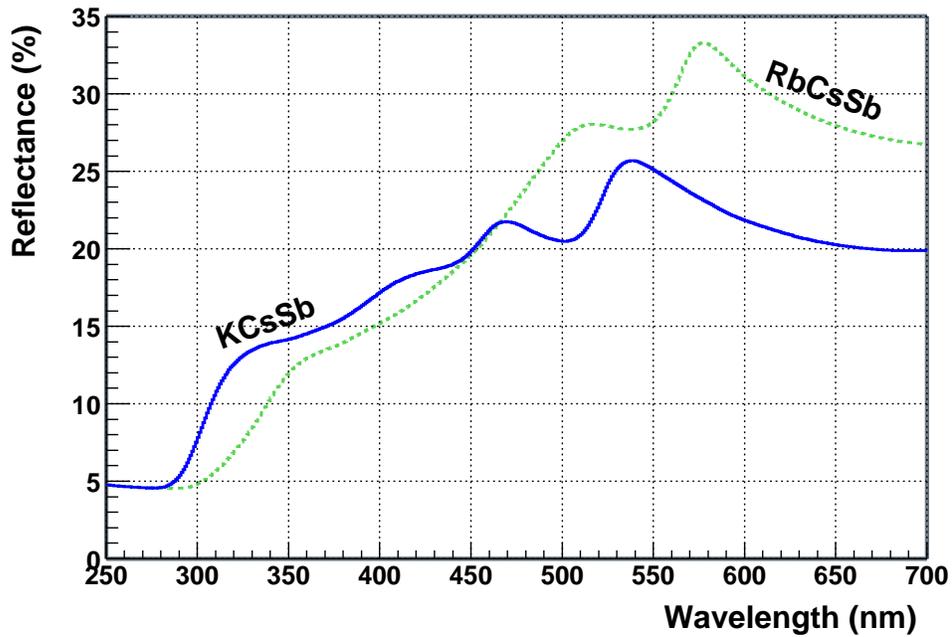}
\end{center}

\caption{\textsl{\small \label{fig:V-W_spectra}Average PMT absolute reflectance
spectra of an ETL 9102B (blue solid curve) and an ETL 9902B (green
dashed curve). The former has a blue-sensitive KCsSb bialkali photocathode,
the latter a RbCsSb green-enhanced bialkali photocathode. The reflectance
is the sum of the Fresnel component from the glass window and the
dominant photocathode contribution. The short-wavelength cut-off is
in both cases due to the absorption from the glass. }}
\end{figure}

At the V-W level of accuracy, the fluctuations within our sample PMTs
due to inhomogeneities of the photocathode are the dominant uncertainty.
Consequently, we report in Fig. \ref{fig:V-W_spectra} the average spectra
from several measurements of both PMTs (relative deviations
of $\sim10\,\%$ are observed).

It is remarkable that both PMTs show a $\sim20\,\%$ reflectance in
the visible region. The same range has also been measured by us in
other bialkali PMTs and seems to be typical of all the bialkali PMTs
\cite{Lang}. Moreover, we found that the spectral shape of the reflectance
is reproduced in other PMTs equipped with the same kind of photocathodes.

\subsection{\label{sub:VASRA}Relative Reflectance at Variable Angles}

\subsubsection*{Experimental Technique}

The reflectance of the test PMTs has also been investigated with a VASRA
accessory (Variable Angle Specular Reflectance Accessory) of our Varian
Cary 400 spectrophotometer. The VASRA allows to measure the specular reflectance
of a sample in the interval $20^{\circ}\leq\theta\leq70^{\circ}$. Unlike the
V-W accessory, the VASRA has no self-referencing capabilities,
therefore any measurement needs to be referenced to a calibrated standard. 

In order to correct for the instrument spectral baseline, we have referenced the VASRA
with the response to a VM2000 \cite{VM2000} reflector
sample, which we have measured to have a nearly flat (at $\sim1\,\%$)
reflectance spectrum at near normal incidence, with
typical $R\sim98\,\%-99\,\%$%
\footnote{This implies an absolute error of $\lesssim2\,\%$, which is however
insignificant, since the absolute scale of the measurement will be
left free (see next paragraph). The only significant systematic error
implicit in this procedure is related to the wavelength dependence
of the VM2000 reflectance. At near-normal incidence, we found $R_{max}-R_{min}\simeq2\,\%$
in the wavelength range used for the analysis \cite{mythesis}. Independent
measurements at $\lambda=430\, nm$ reported in the same reference
give an indication that $R\gtrsim98\,\%$ at all incidence angles.%
} \cite{mythesis}. \textsl{\small }The measurements were limited to
$\lambda\geq400\, nm$, because the VM2000 used as reference has reflectance
cut-off at $\lambda\sim390\, nm$\textsl{\small .} A polarizer is
employed to select the \emph{p} and \emph{s} waves before reflection
from the sample, and a depolarizer is mounted at the end of the optical
path to avoid biases related to the sensitivity of the spectrophotometer
light sensor to the polarization state of the outgoing light.

Three angles of incidence are considered for the final analysis: $\theta=45^{\circ},\:55^{\circ},\:65^{\circ}$.
This angular range is chosen because the strongest constraints in
the determination of the photocathode optical parameters are expected
from measurements at high angles of incidence. Furthermore, the three
selected angles are around the Brewster's angle for Fresnel reflection
off the glass, so that the PMT reflectance of the parallel polarization
is largely dominated by the photocathode contribution.

\subsubsection*{Results and Discussion}

\begin{figure} 
\begin{center}
 \includegraphics[width=1.0\columnwidth,
  keepaspectratio]{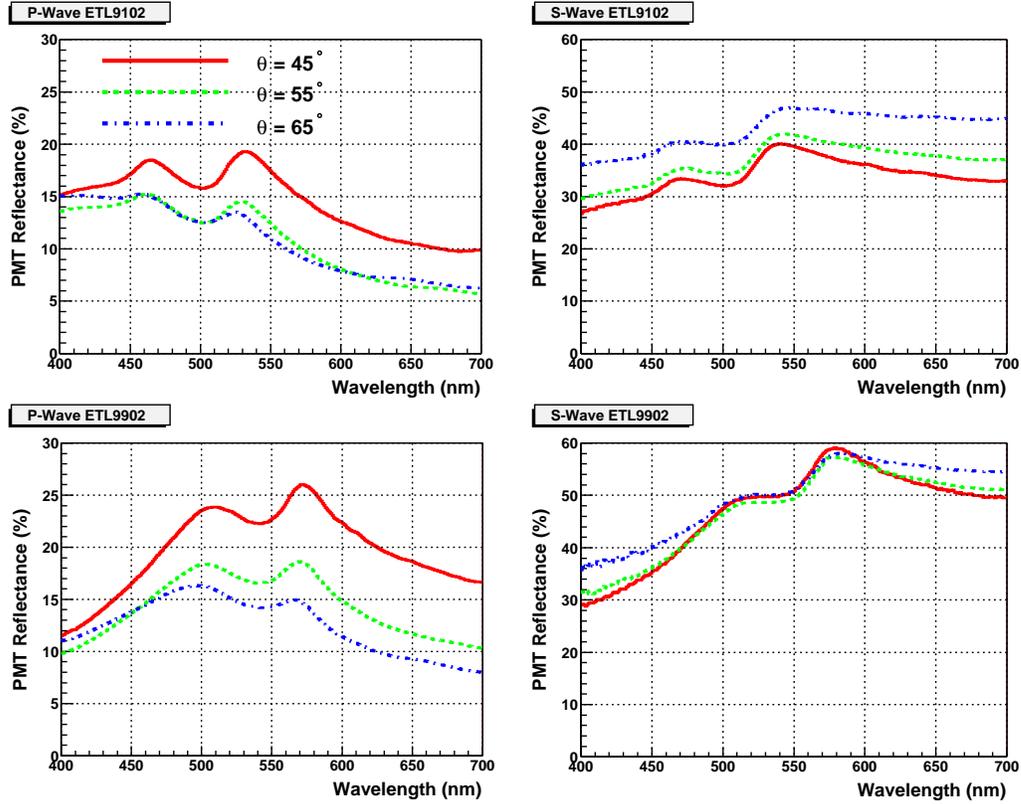}
\end{center}

\caption{\textsl{\small \label{fig:VASRA_gen} Reflectance spectra of
the PMTs ETL 9102B (top) and ETL 9902B (bottom) for parallel (left)
and perpendicular (right) polarized light, recorded with a Varian
Cary400 spectrophotometer by using the VASRA accessory. Ordinates
are given in percent of the incident light, but the uncertainty in
the absolute normalization of the measurements is large (see text).
The three shown spectra per PMT and polarization state correspond
to different angles of incidence: $\theta=45^{\circ}\textrm{(red)},\textrm{ }55^{\circ}\textrm{(green dashed), }65^{\circ}\textrm{(blue dot-dashed)}$.}}
\end{figure}

Fig. \ref{fig:VASRA_gen} displays the average reflectance spectra,
where the mean values of a set of $\geq4$ repeated scans are shown.
We found that the shape of the curves are very reproducible, however
the absolute scale of the measurement is not (relative deviations
of up to $\sim20\,\%$ are observed). The reason is that the alignment
of the VASRA sample holder can change slightly for different scans,
resulting in a variation of the collection efficiency for the reflected 
light. For this reason, the VASRA measurement will be used for 
spectral shape data only, while the absolute normalization of all 
curves is considered as a free parameter of the fit function 
(to be discussed in Sec. \ref{sec:Data-analysis}).

\subsection{\label{sub:Ellipsometric-measurements}Ellipsometric Measurements}

\subsubsection*{Experimental Technique}

The ellipsometric measurements have been performed with a \emph{M-44
Vis J.A. Woollam} spectral ellipsometer. The instrument allows the
simultaneous measurement of 44 discrete wavelengths in the range $[419\, nm-742\, nm]$.
The angle of incidence on the sample was $\theta=60.75^{\circ}\pm0.01^{\circ}$,
measured by ellipsometry itself on a reference $SiO_{2}$ film. Light
was linearly polarized and the polarization state of light upon reflection
was analyzed by decomposing the elliptical polarization into
the parameters $\Psi\textrm{ and }\Delta$ of Eq. \ref{eq:ellipsometry}.
The thickness of the PMT glass window was sufficient to allow a clear
separation of the beams reflected off the glass and the photocathode,
so that it was possible to select only the latter light to the analyzer. 
This selection facilitates the succeeding interpretation
of data. For both PMTs three ellipsometric scans have been performed,
corresponding to the illumination of different spots of the photocathode.

\subsubsection*{Results and Discussion}

\begin{figure} 
\begin{center}
 \includegraphics[width=1.0\columnwidth,
  keepaspectratio]{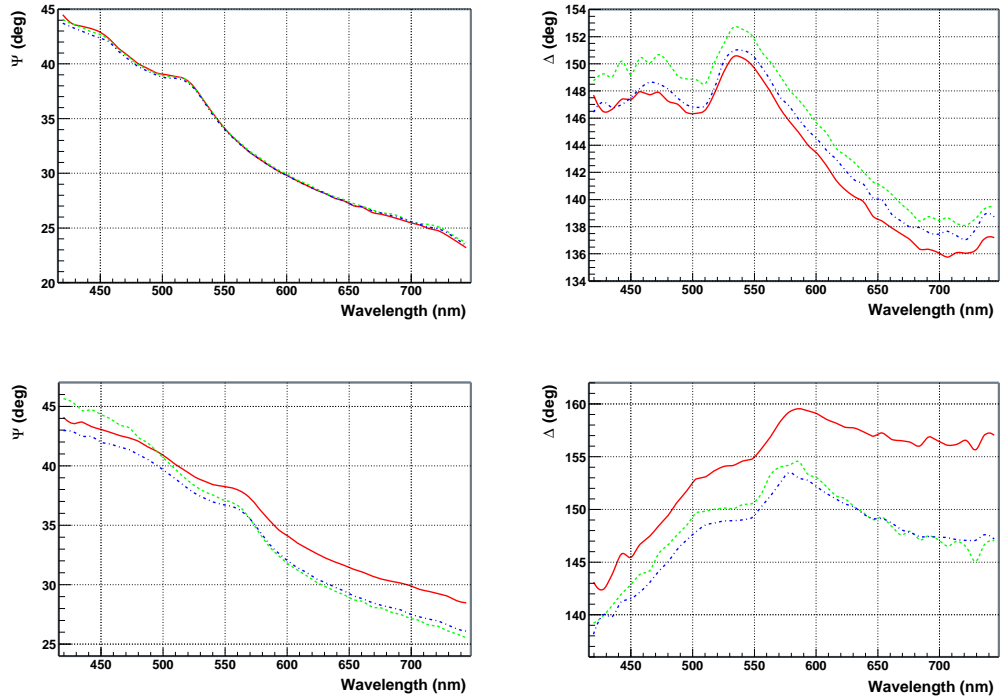}
\end{center}

\caption{\textsl{\small \label{fig:ellips_data}Ellipsometric measurements
on a ETL 9102B (top) and a ETL 9902B (bottom) PMT. The graphics
show the wavelength spectra of the $\Psi$ and $\Delta$ functions.
The three curves in each frame correspond to different illuminated spots 
on the PMTs (red solid: center; green dashed and blue dashed-dotted: close 
to the edge, at $\sim90^{\circ}$ from each other). The plots refer to 
the measurement of the light reflected from the back-plane of the PMT window, 
where the photocathode layer is situated. The instrument was set at an ellipsometric
angle of $\theta=60.75^{\circ}$.}}
\end{figure}

The results are shown in Fig. \ref{fig:ellips_data}. It can be
seen that for the PMT ETL 9102B the three scans of the $\Psi(\lambda)$
function are in excellent agreement with each other, while the $\Delta(\lambda)$
spectra measured in different positions are similar, but shifted by
nearly constant phases. This is likely to be an artifact introduced
by stress-birefringence of the PMT glass window. This stress is different 
in different positions, and this can explain the observed
phase shifts in the three measurements of $\Delta(\lambda)$.
Differences are also observed in the $\Psi(\lambda)$ scans of the
ETL 9902B PMT. It is likely that this effect is due to a inhomogeneous
photocathode thickness\footnote{The process of photocathode growth through evaporation can lead to
some inhomogeneities in the layer thickness. This is for example
observed and studied in \cite{Lay,Lang}.}.

\section{\label{sec:Data-analysis}Data Analysis}

In Sec. \ref{sec:Theory-and-methodology} the equations describing
the optical model of a thin absorbing film have been introduced and
in Sec. \ref{sec:Opt-meas-photocath} all the optical measurements
performed on two sample PMTs have been reported. The purpose of data
analysis is to combine theory and experiments to derive the unknowns
of the problems. Those unknowns are:

\begin{enumerate}
\item The photocathode refractive index, $n^{*}(\lambda)=n(\lambda)+ik(\lambda)$
\item The photocathode thickness, $d$
\end{enumerate}
For simplicity, it will be assumed that the glass refractive index
is a known function: \begin{equation}
n_{glass}=1.472+3760/\lambda^{2}\label{eq:cauchy_glass}\end{equation}
Eq. \ref{eq:cauchy_glass} is the Cauchy dispersion law with the
typical parameters of the borosilicate glass employed for the PMT
envelope%
\footnote{The refractive index of the borosilicate glass is measured by many
manufacturers to be in the range
$1.48\lesssim n\lesssim1.50$ in the visible spectrum. Eq. \ref{eq:cauchy_glass} 
gives $n(380\,nm)\simeq 1.50$, $n(680\,nm)\simeq 1.48$ and $n(442\,nm)\simeq 1.49$. 
The latter value has been directly measured by M\&T \cite{Moorhead}.  It has been
also verified that fixing $n_{glass}=c$, where \emph{c} is any constant in the
above range, leads to similar results in the analysis.}.

We have written a computer program implementing Eqs. \ref{eq:R&T_amplitudes_thin_film}
to \ref{eq:Fresnel_s,p}, which is used to perform a global fit
to the experimental data based on a least square minimization. The
program uses the package MINUIT of the CERN software libraries to
perform such minimization. The global parameter $d$ is easily implemented,
while real and imaginary part of the photocathode refractive index
are continuous functions of the wavelength. They are
implemented in the global fit as a table of parameters at discrete
wavelengths, from $380\, nm$ to $680\, nm$ with steps of $15\, nm$,
and then interpolated at each data-point wavelength by using a cubic
\emph{spline} function. As mentioned in Sec. \ref{sec:Opt-meas-photocath},
some of our measurements suffer from systematics implying a partial
loss of information. This forces the introduction of new free parameters
in the fit:

\begin{enumerate}
\item Six normalization factors for the VASRA reflectance measurements, 
expressing the systematic uncertainty in the absolute scale of the scans
\item One constant phase-shift for the average $\Delta$ ellipsometric spectrum,
accounting for the effect of the glass birefringence
\end{enumerate}
The fit is performed in the wavelength range $[380\, nm,\,680\, nm]$%
\footnote{In the interval $[380\, nm,\,400\, nm]$ only the V-W data is available.
In $[400\, nm,\,420\, nm]$ V-W plus VASRA. Above $420\, nm$ the
full data-set. The degeneracy in the $[380\, nm,\,400\, nm]$ interval is broken by 
requiring the continuity of the optical functions (the photocathode thickness is common to
all wavelengths).}. 

The function \emph{FCN} minimized by MINUIT is expressed as the sum
of 4 contributions: $FCN=FCN_{V-W}+FCN_{VASRA}+FCN_{\Psi}+FCN_{\Delta}$.
Each one, $FCN_{l}$, is defined as:\begin{equation}
FCN_{l}(p_{j})=\frac{1}{n}\sum_{i=1}^{n}\left[y_{li}-m_{l}(\lambda_{i},\textrm{ }p_{j})\right]^{2}\label{eq:FCN}\end{equation}
 where $y_{li}$ is the value of the $l^{th}$ optical measurement at the wavelength
$\lambda_{i}$ and $m_{l}(\lambda_{i},\textrm{ }p_{j})$ is the prediction
of the theoretical model, for a particular choice of the parameters
set $\{ p_{j}\}$. Eq. \ref{eq:FCN} is normalized to the total
number of data-points $n$ to have each measurement contribute the 
same statistical weight in the global fit.

\section{\label{sec:Results}Results}

\subsection{Preliminary Remarks and Analysis Strategy}

The inversion of our experimental data by using
the described fitting procedure does not lead to a unique
solution. We find that our measurements are sufficient to constrain the spectral shape
of the optical constants, however there are strong correlations in the fit between
\emph{n}, \emph{k} and \emph{d}, so that the FCN function does not show a well defined
minimum. The reason is that our measurements are not really
``orthogonal'' and because of the systematics discussed in Sec.
\ref{sec:Opt-meas-photocath}, the VASRA and ellipsometric data provide 
only partial information.
Furthermore, measurements in air give weaker constraints than measurements in
higher refractive index media, as the model is highly degenerate in the optical 
parameters for this case (see Sec. \ref{sub:Experimental-Approaches}). 

The values of the photocathode complex refractive index measured by M\&T, Lang and Lay at
$\lambda=442\, nm$  provide the missing constraint to break the residual degeneracy
at all wavelengths. In fact, solutions can be found consistent with those
previous determinations, however provided that the ellipsometric $\Psi(\lambda)$ data
are corrected by introducing a free angular shift. Since the physical meaning of
such a parameter is less clear than for the case of the expected phase-shift 
of the $\Delta(\lambda)$ function, in this paper we present an analysis 
of our data in which $\Psi(\lambda)$ is not included in the global 
fit. The consistency of the $\Psi(\lambda)$ spectrum with the other 
data and the optical model is discussed in the next section.

The analysis for the \emph{KCsSb} photocathode is carried out by requiring the 
solution to be consistent with the optical constants measured by M\&T at $442\, nm$, and
later confirmed by Lang and Lay. This means that we fix $n=2.7\pm0.2$, $k=1.5\pm0.2$ 
at $\lambda=440\, nm$ (the closest wavelength to $442\, nm$ 
in our parameters table). The only published measurements on \emph{RbCsSb} 
photocathodes using the M\&T technique are reported in \cite{Lang}. A large scatter 
is observed, however the two samples manufactured by ETL and expected to be very
similar to our ETL 9902B give consistent results. We therefore
fix the optical constants at $\lambda=440\, nm$ to the average value
of these PMTs: $n=2.5\pm0.2$, $k=1.35\pm0.1$. For either photocathode
the quoted errors are $\sim\pm2\sigma$.

Fixing the optical constants at the given wavelength in the global
fit (where the $\Psi(\lambda)$ scan is excluded) breaks the degeneracy
and leads to unique solutions for the photocathode thickness \emph{d}
and the spectra $n(\lambda)$ and $k(\lambda)$. The uncertainty associated
with the error of the optical parameters at $440\, nm$ is accounted
for by calculating an ``upper'' solution with fixed $(n,\, k)_{KCsSb}=(2.9,\,1.7)$
and $(n,\, k)_{RbCsSb}=(2.7,\,1.45)$, and a ``lower'' solution
for $(n,k)=(2.5,\,1.3)_{KCsSb}$ and $(n,k)_{RbCsSb}=(2.3,\,1.25)$.
The band between these two solutions defines our $2\sigma$ confidence
interval for real and imaginary part of the complex refractive index
at all wavelengths, and as well for the photocathode thickness. This choice
of estimating the errors is justified by the fact that in M\&T's and
Lang's fit to data, as well as in ours, \emph{n} and \emph{k} are
strongly correlated. We estimate that the additional uncertainty coming from 
the fit of our data is small compared to the one associated with the
error of the fixed optical parameters.

\subsection{KCsSb Bialkali Photocathode}

\begin{figure} 
\begin{center}
 \includegraphics[height=15cm,
  keepaspectratio]{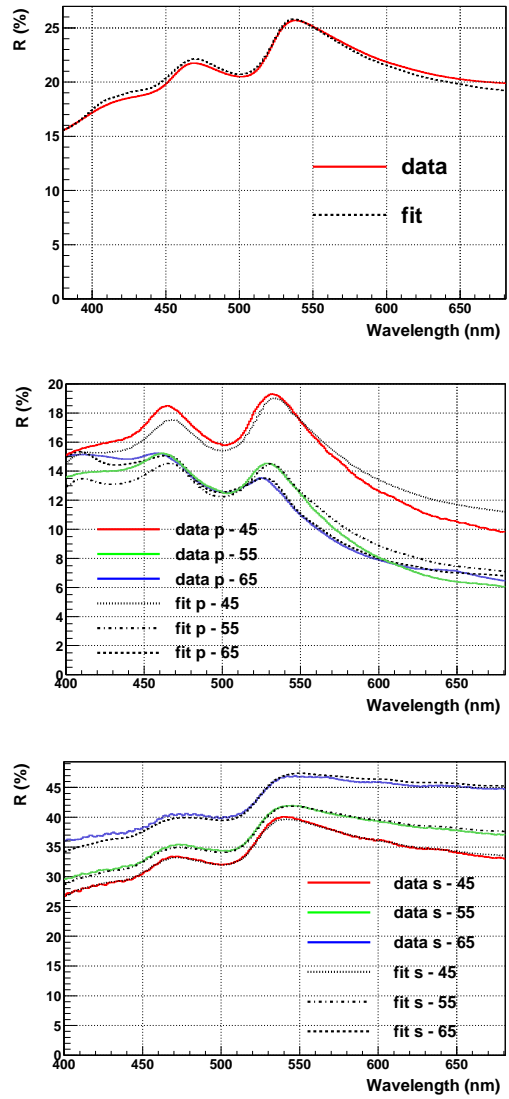}
\end{center}

\caption{\label{fig:ex_globalfit_1}\textsl{\small Best global fit to
the optical measurements of the ETL 9102B sample PMT, for the case
of optical parameters fixed to $n(440\, nm)=2.7$, $k(440\, nm)=1.5$. }}

\textbf{\textsl{\small Top}}\textsl{\small : V-W absolute reflectance
spectrum.}{\small \par}

\textbf{\textsl{\small Center}}\textsl{\small : VASRA reflectance,
parallel polarization.}{\small \par}

\textbf{\textsl{\small Bottom}}\textsl{\small : VASRA reflectance,
perpendicular polarization}{\small \par}

\textsl{\small The measurements are displayed in solid lines
(see Figs. \ref{fig:V-W_spectra} and \ref{fig:VASRA_gen}),
the fits in dashed (dashed-dotted and dotted) lines. }{\small \par}

\textsl{\small The fits obtained after fixing $(n,\, k)$ to the other
considered values are very similar to the one shown here.}
\end{figure}

\begin{figure} 
\begin{center}
 \includegraphics[width=0.80\columnwidth]{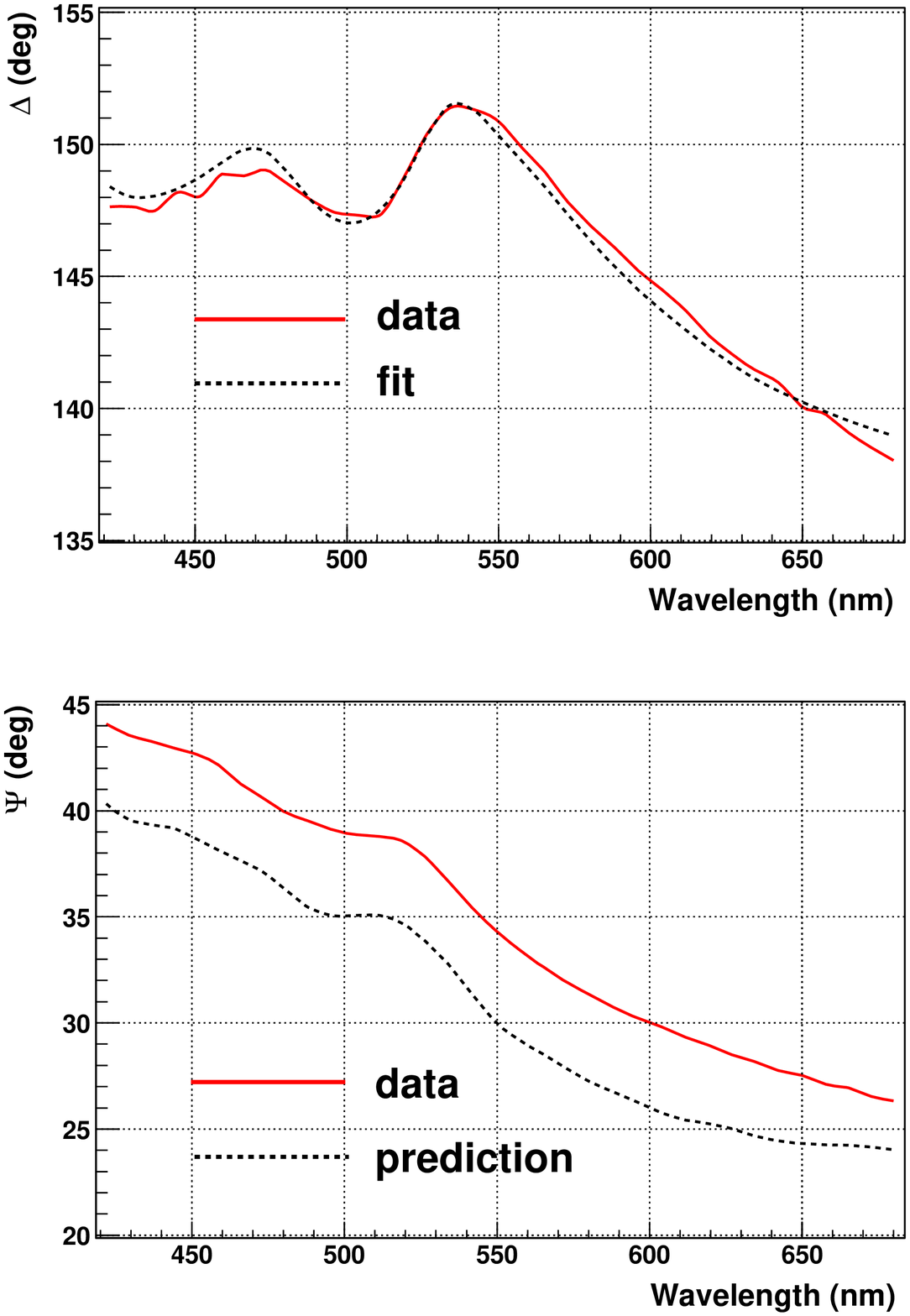}
\end{center}

\caption{\textbf{\textsl{\small \label{fig:ex_globalfit_2}}}}

\textbf{\textsl{\small Top}}\textsl{\small : best fit (black dashed)
to the measured average ellipsometric $\Delta$ spectrum of the ETL
9102B (red solid), for the case of optical parameters fixed to $n(440\, nm)=2.7$,
$k(440\, nm)=1.5$. }{\small \par}

\textbf{\textsl{\small Bottom}}\textsl{\small : Measured average $\Psi$
spectrum (red solid, not included in the global fit) and relevant
prediction of the model giving the best global fit to V-W, VASRA and
$\Delta$ data (black dashed).}{\small \par}

\textsl{\small The fits obtained after fixing $(n,\, k)$ to the other
considered values are very similar to the one shown here.}
\end{figure}

\begin{figure} 
\begin{center}
 \includegraphics[width=1.0\columnwidth,
  keepaspectratio]{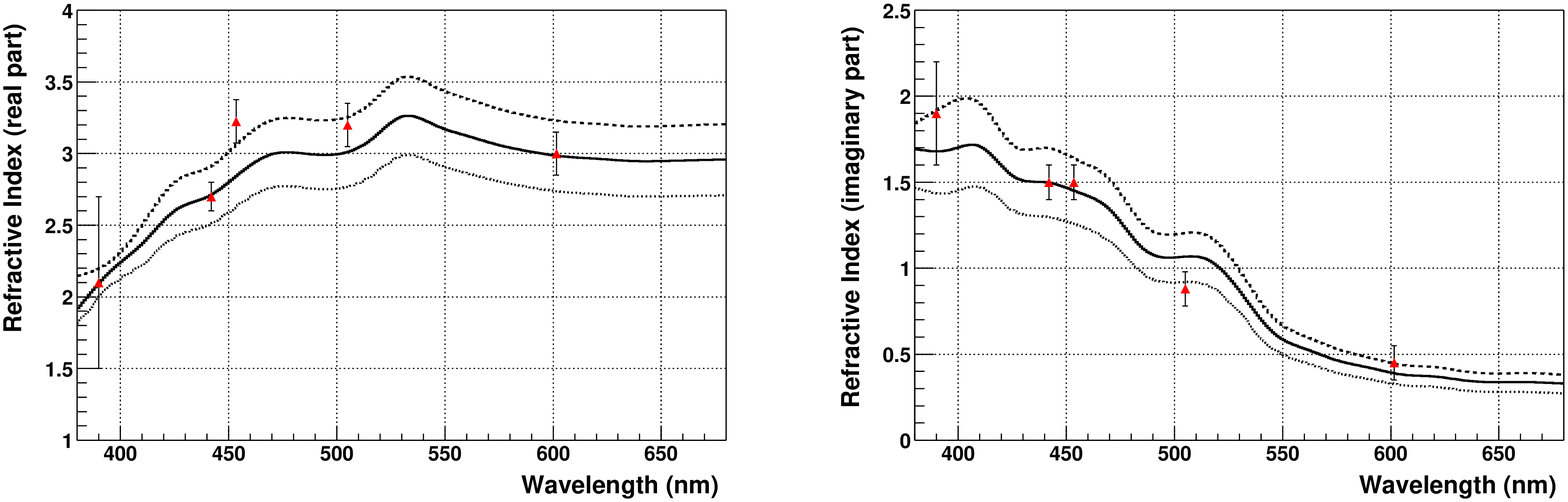}
\end{center}

\caption{\label{fig:bluebialkali_N-K}\textsl{\small Best fit spectra
and $2\sigma$ allowed bands for real (left) and imaginary (right)
part of the complex refractive index of a KCsSb photocathode. The
solid curves correspond to the global fit constrained by fixing n
and k to M\&T's best values at $\lambda=440\, nm$: $n=2.7$, $k=1.5$.
The dashed and dotted curves are the results of the fit after fixing
the optical constants at the above wavelength $2\sigma$ away from
M\&T's best value. The markers show M\&T's analysis of their own data
at $\lambda=442\, nm$ and their re-analysis of previous measurements
from Timan \cite{Moorhead,Timan}. Error bars are $\pm1\sigma$.}}
\end{figure}

An example of global fit to the optical measurements performed on
the PMT ETL 9102B is shown in Figs. \ref{fig:ex_globalfit_1}
and \ref{fig:ex_globalfit_2}, while Fig. \ref{fig:bluebialkali_N-K}
shows the best fit solutions for the complex refractive index and
the $2\sigma$ allowed bands. In Table \ref{tab:blue_fitpar}
the other parameters of the fit are reported.

Fig. \ref{fig:ex_globalfit_2} also shows the prediction of the
best fit model for the $\Psi(\lambda)$ ellipsometric function, 
not used for the global fit. 
It is remarkable that the predicted spectral shape is in excellent
agreement with the measurement. Data and model simply differ by a
constant angular shift, which is probably due to an unimplemented
systematic effect related to the glass birefringence. An additional
free offset might be introduced to include the $\Psi(\lambda)$ data
in the global fit. However the figure shows that this information
would simply be redundant.

The results in Fig. \ref{fig:bluebialkali_N-K} show that the
spectral shape of the real and imaginary part of the photocathode
refractive index is well constrained by the data, while the absolute
scale depends on the assumptions of the optical constants
fixed during the fit. Both \emph{n} and \emph{k} anti-correlate with
the photocathode thickness (Table \ref{tab:blue_fitpar}), 
as was observed also by M\&T in their analysis.

Fig. \ref{fig:bluebialkali_N-K} also shows a graphical comparison
of our results with those of M\&T, including a re-analysis that M\&T
propose of previous measurements by Timan at discrete wavelengths
in the visible region \cite{Moorhead,Timan}. Our results are in good
agreement with M\&T's analysis of Timan, both for the real and imaginary part
of the complex refractive index. 

The estimation of the photocathode thickness (Table \ref{tab:blue_fitpar})
is consistent with the typical values found by M\&T, Lang and Lay, 
and expected by the PMT manufacturer.

\subsection{RbCsSb Bialkali Photocathode}

\begin{figure} 
\begin{center}
 \includegraphics[width=1.0\columnwidth,
  keepaspectratio]{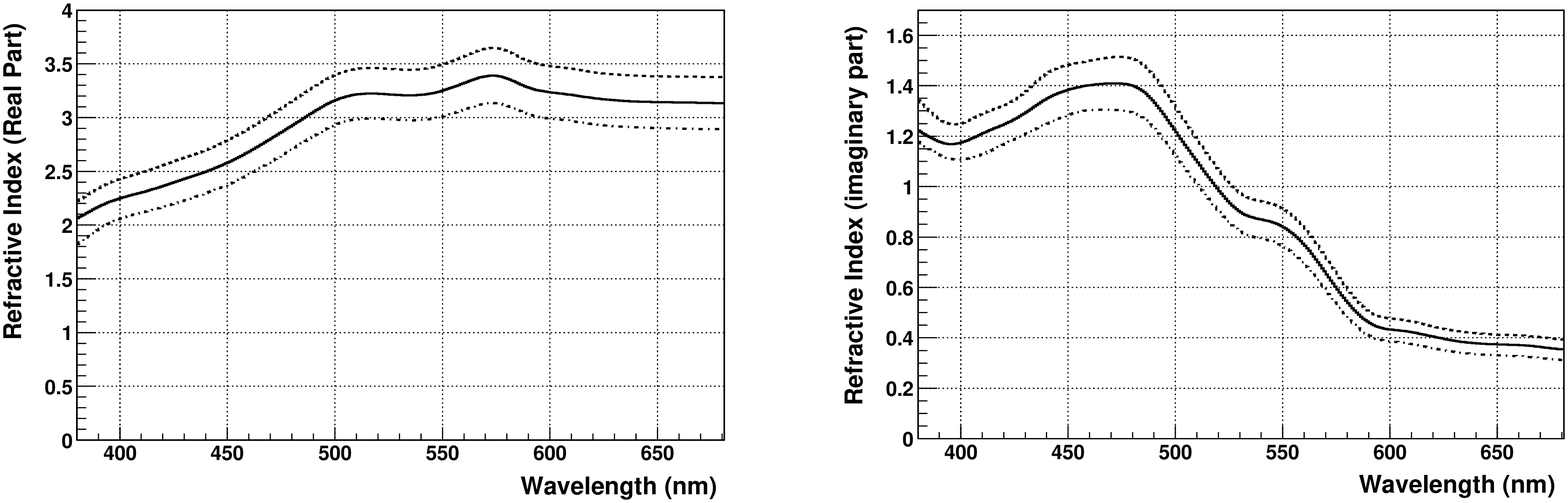}
\end{center}

\caption{\textsl{\small \label{fig:greenbialkali_N-K}Best fit spectra
and $2\sigma$ allowed bands for real (left) and imaginary (right)
part of the complex refractive index of a RbCsSb photocathode.The
solid curves correspond to the global fit constrained by fixing n
and k to Lang's best values at $\lambda=440\, nm$: $n=2.5$, $k=1.35$.
The dashed and dotted curves are the results of the fit after fixing
the optical constants at the above wavelength $2\sigma$ away from
Lang's best value.}}
\end{figure}

We have carried out a similar analysis for the optical measurements
on the PMT ETL 9902B (\emph{RbCsSb} bialkali). The estimated real
and complex part of the refractive index as a function of the wavelength
are plotted in Fig. \ref{fig:greenbialkali_N-K}.

The solutions show similar features to the case of the \emph{KCsSb}:
the spectral shapes are well constrained by data and shifting the
values at $440\, nm$ fixed during the fit produces a corresponding
shift of the entire spectra, which is compensated by an anti-correlated
change in the photocathode thickness (Table \ref{tab:green_fitpar}).

\section{Discussion and Applications}

In the previous section it has been shown that our measurements, in
conjunction with M\&T's (\emph{KCsSb}) and Lang's (\emph{RbCsSb})
at $\lambda=442\, nm$, lead to unique solutions for the optical constants
of the investigated photocathodes in the full visible range. In the
following sections we report the optical performances of bialkali
PMTs predicted by our model for some specific cases. We also try to
probe the degree of predictability of our model and to evaluate the
uncertainties originating from the ambiguity left in the problem.
In all the calculations it will be assumed that light impinging on
the PMT is unpolarized.

It should be noted that even with a precise determination of the photocathode
optical parameters in one PMT, the problem of predicting the behavior
of any other similar PMT would still remain to some extent indeterminate,
because the thickness of the photocathode can vary up to a factor $\sim2$
from one PMT to another, as shown by Lang. Therefore we point out that the predictions
and relevant uncertainties presented in the next sections refer specifically to the
two measured sample PMTs. The model can be extended to any generic \emph{KCsSb} and
\emph{RbCsSb} PMT, however with a somewhat larger uncertainty. We do not discuss here the
effect of the dependence of the PMT optical properties on the photocathode thickness. 
This issue is examined in \cite{Moorhead,Lay}.

Several cases may be considered as for the coupling of the PMT to
the medium where the light originates. We will focus on the case of
a \emph{KCsSb} PMT optically coupled to a scintillator. For simplicity,
it will be assumed $n_{medium}=1.48$ at all wavelength, so that the
medium refractive index is slightly lower than that of the PMT window glass
at all wavelengths in the visible.

\subsection{Wavelength Dependence of the Reflectance}

\begin{figure} 
\begin{center}
 \includegraphics[width=1.0\columnwidth,
  keepaspectratio]{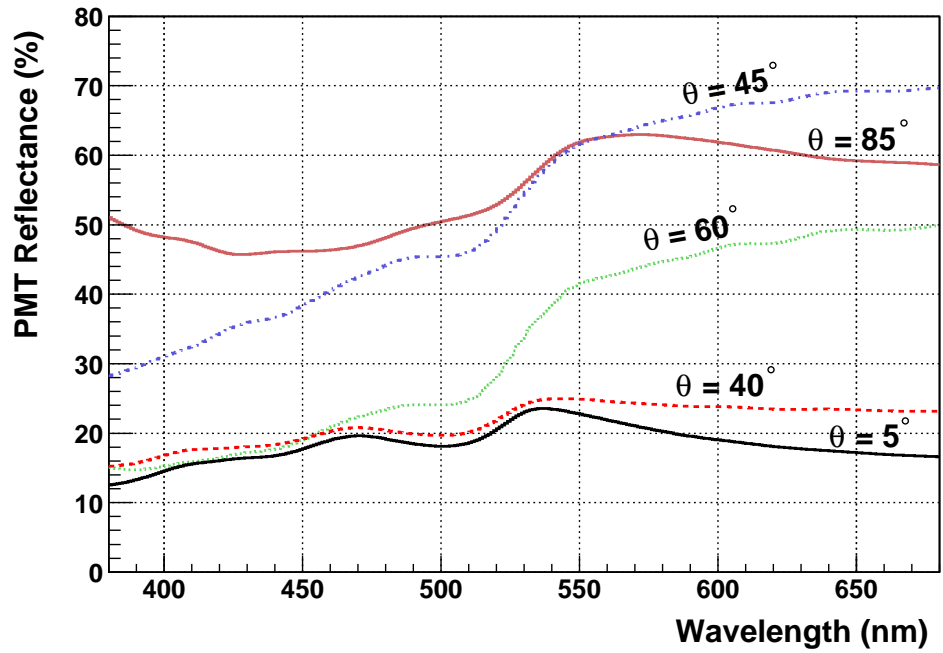}
\end{center}

\caption{\textsl{\small \label{fig:blue_R_model} PMT reflectance as a
function of the wavelength at some angles of incidence, calculated
for the sample ETL 9102B in a medium with $n=1.48$, choosing the
central solution shown in Fig. \ref{fig:bluebialkali_N-K}.}}
\end{figure}

Fig. \ref{fig:blue_R_model} reports the wavelength dependence
of the PMT reflectance at various angles, calculated for our sample
ETL 9102B in (or coupled to) a medium with $n=1.48$. 

\begin{figure} 
\begin{center}
 \includegraphics[width=1.0\columnwidth,
  keepaspectratio]{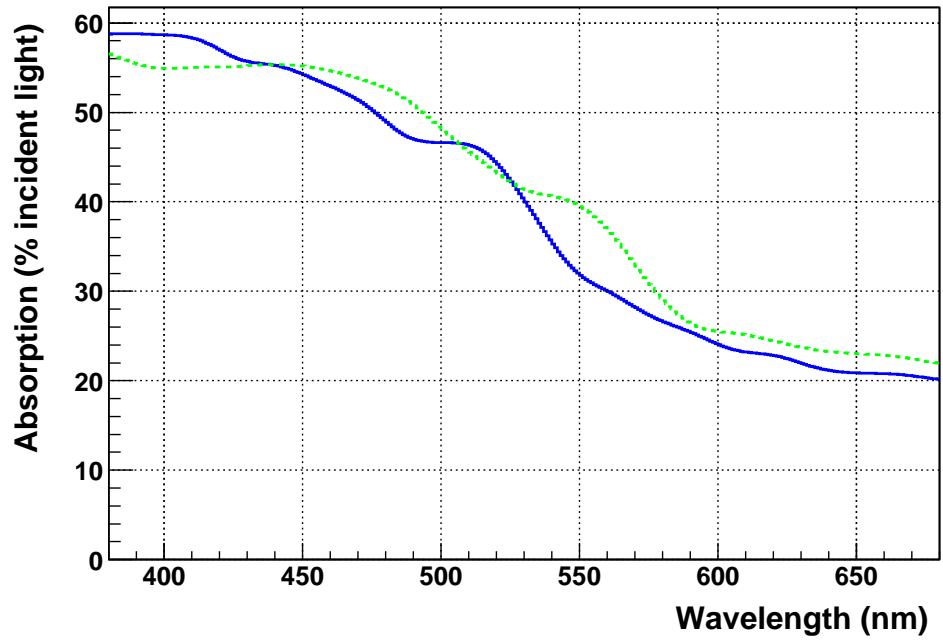}
\end{center}

\caption{\textsl{\small \label{fig:abs(lam)_air}Photocathode absorption
calculated for PMTs in air and light impinging with $\theta=5^{\circ}$.
The blue-solid curve refers to the ETL 9102B sample, the green-dashed
to the ETL 9902B. The central solutions in Figs \ref{fig:bluebialkali_N-K},
\ref{fig:greenbialkali_N-K} and Tables \ref{tab:blue_fitpar},
\ref{tab:green_fitpar} are chosen.}}
\end{figure}

At near normal incidence the spectrum is very similar to the one measured
in air (cf. Fig. \ref{fig:V-W}) and does not change much until
$\theta$ approaches $\theta_{c}$ ($\simeq42^{\circ})$. Above the
critical angle the spectrum behaves differently, with higher reflectance
at longer wavelengths: since transmission is suppressed for $\theta>\theta_{c}$
and absorption decreases with $\lambda$ (to discuss in Sec. \ref{sub:QE_qualitative}, 
cf. Fig. \ref{fig:abs(lam)_air}),
the reflectance must necessarily compensate, to preserve $A+R+T=1$.

The predictions for the \emph{RbCsSb} photocathode
are qualitatively very similar. The only visible difference is that
the main features of the spectra (peaks below $\theta_{c}$, slope
change above) appear $\sim40\, nm$ shifted to longer wavelengths,
coherently with all the optical measurements. 

\begin{figure} 
\begin{center}
 \includegraphics[width=1.0\columnwidth,
  keepaspectratio]{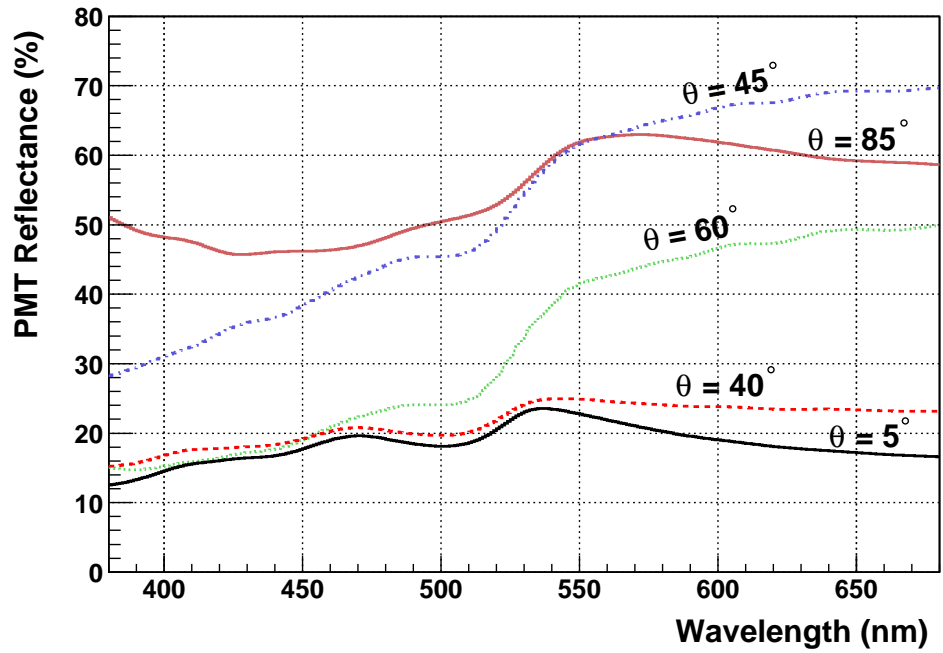}
\end{center}

\caption{\textsl{\small \label{fig:blue_R_model_15-25}Comparison between
the PMT reflectance spectra (ETL 9102B) calculated using the ``upper''
(dashed-dotted) and ``lower'' (dashed) solutions of Fig. \ref{fig:bluebialkali_N-K}
and the corresponding thickness in Table \ref{tab:blue_fitpar}.
The two curves for $\theta=5^{\circ}$ are not distinguishable in
the figure.}}
\end{figure}

In Fig. \ref{fig:blue_R_model_15-25} it is shown how the indetermination
in our model affects the predictions:
 for $\theta<\theta_{c}$ all the solutions in the allowed bands give
nearly the same results, whereas at higher angles some differences
are observed. This is related to the fact that the optics of a PMT
in liquid below the critical angle does not differ very much from
that of a PMT in air (after taking into account the Fresnel reflection
from the glass). The selected solutions are those that fit best the
measurements in air, consequently they also make very similar predictions
for a PMT in liquid at $\theta<\theta_{c}$. On the other hand, also
above the critical angle the difference in the predictions of the
allowed solutions is $<10\,\%$.

\subsection{Angular Dependence of Reflectance and Absorption}

\begin{figure} 
\begin{center}\includegraphics[width=1.0\columnwidth,
  keepaspectratio]{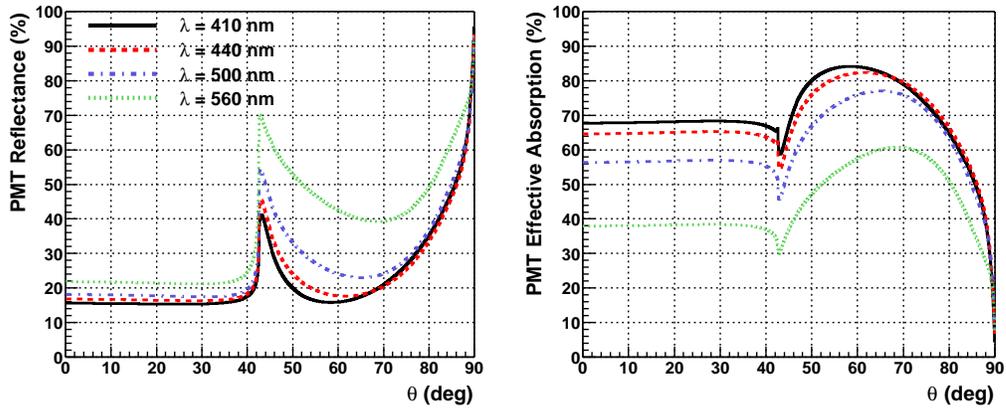}\end{center}

\caption{\textsl{\small \label{fig:RA_theta-lam}Predicted angular dependence
of reflectance and effective absorption (see footnote \ref{footnote:total-absorption}
for the definition), for the investigated ETL 9102B PMT in optical
contact with a scintillator ($n=1.48$), at 4 representative wavelengths. }}
\end{figure}

In Fig. \ref{fig:RA_theta-lam} the angular dependence of the
PMT reflectance and effective absorption%
\footnote{\label{footnote:total-absorption}The effective absorption includes
also the fraction of transmitted light that is back reflected onto
the photocathode by the aluminized internal surface of the PMT, and is 
eventually absorbed. For simplicity we do not simulate the optics inside the PMT
and assume that on average a fraction $\sim0.7$ of the transmitted light ($T$) hits
the photocathode again. The probability for absorption is given by the same optical
model described in Sec. \ref{sec:Theory-and-methodology}, simply
by exchanging the order of the refractive indices, going from vacuum
to scintillator. Absorption from the photocathode side is a very flat
function of $\theta$. Using the central solution model for the PMT
ETL 9102B, we find: $A(410\, nm)\sim0.44\textrm{, }A(440\, nm)\sim0.41\textrm{, }A(500\, nm)\sim0.35$,
$A(560\, nm)\sim0.22$. These values are multiplied by $0.7T$ and
added to the absorption from the scintillator side to obtain the effective
absorption function. The effect of this correction is to increase
absorption below the critical angle, resulting in a higher QE. As
a by-product, the PMT efficiency has a more uniform angular dependence. %
} is shown for 4 representative wavelengths. In all cases the reflectance
has a plateau below $\theta_{c}$, then a sharp peak at the critical
angle, followed by an extended angular region where \emph{R} has similar
or higher values than at near normal incidence; last \emph{R} has
a fast increase to $100\,\%$ for $\theta\rightarrow90^{\circ}$. Similarly
the effective absorption has a flat angular dependence for $\theta<\theta_{c}$,
a dip at the critical angle, followed by a broad peak and it then
drops to zero, since $A+R=100\,\%$ ($T=0$ for $\theta>\theta_{c}$).
At different wavelengths the branching between reflection and absorption
changes, as well as the shape of the peak around the critical angle.

\begin{figure} 
\begin{center}
 \includegraphics[width=1.0\columnwidth,
  keepaspectratio]{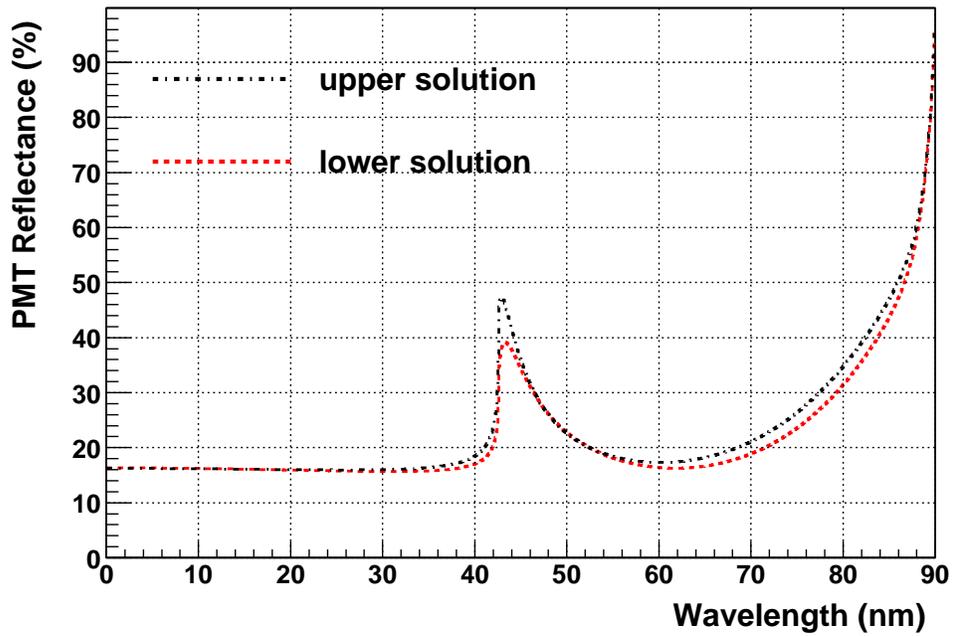}
\end{center}

\caption{\label{fig:R(theta)_models}\textsl{\small Reflectance calculated
for the investigated PMT ETL 9102B operating in optical contact with
a scintillator ($n=1.48$), at the wavelength $\lambda=425\, nm$.
The predictions for the limit ``upper'' and ``lower'' solutions
of the allowed bands are shown. }}
\end{figure}

In Fig. \ref{fig:R(theta)_models} the impact on $R(\theta)$ of the uncertainty
of the optical parameters is studied, for the PMT ETL 9102B at $\lambda=425\, nm$.
It is again found that all the allowed solutions give the same predictions for
$\theta<\theta_{c}$, whereas they differ slightly from each other
at larger angles of incidence. The strongest departures ($\lesssim20\,\%$)
are observed in a narrow angular range around $\theta_{c}$ and for
$65^{\circ}\lesssim\theta\lesssim80^{\circ}$.

The analysis of the \emph{RbCsSb} photocathode leads to very similar
predictions. In case $n_{medium}>n_{glas}$ the same qualitative features
are observed, however, the reflectance (absorption) goes to $100\,\%$
($0\,\%$) below $90^{\circ}$, corresponding to the condition for
total internal reflection at the medium-glass interface.

\subsection{Angular Dependence of the Quantum Efficiency}

\begin{figure} 
\begin{center}
 \includegraphics[width=1.0\columnwidth,
  keepaspectratio]{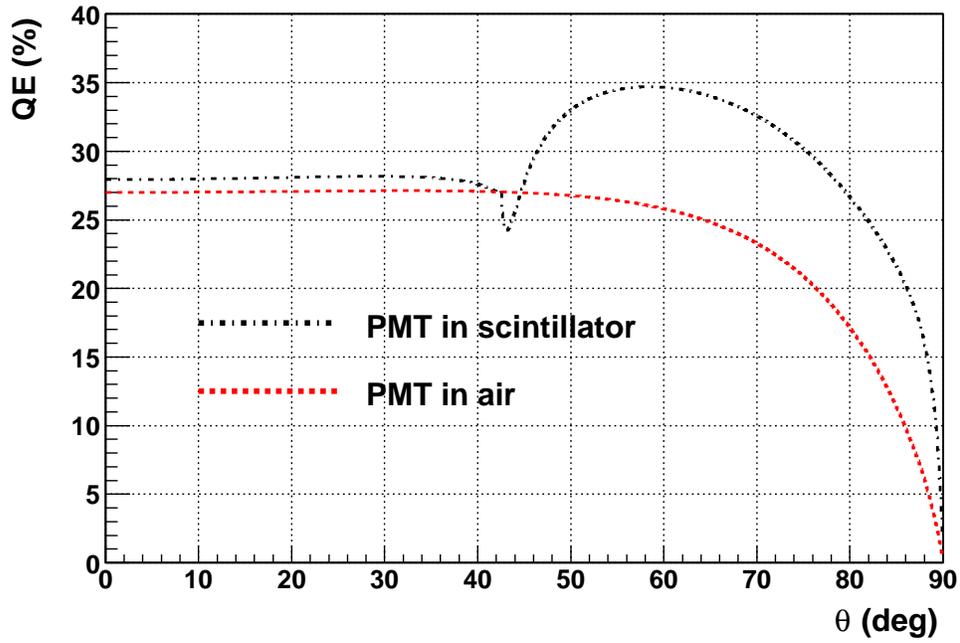}
\end{center}

\caption{\label{fig:QE(theta)_scint-air}\textsl{\small Predicted angular
dependence of the QE, for the case of the ETL9102B PMT in scintillator
($n=1.48$, blue dashed curve) and in air ($n\simeq1$, red dashed-dotted
curve). The optical parameters corresponding to $\lambda=410\, nm$
are used and the plotted curves are obtained by rescaling the effective
absorption, assuming $QE(410\, nm)=27\,\%$ in air, for $\theta=0^{\circ}$. }}
\end{figure}

Assuming that the conversion factor $P_{conv}(\lambda)$ in Eq. \ref{eq:QE}
does not depend on $\theta$ (see \cite{Lay}), the angular dependence
of the PMT QE is the same as that of the effective absorption. As
a consequence, the efficiency of a PMT can be predicted for any $\theta$
and $\lambda$ by rescaling the calculated effective
absorption at $\theta$ with $P_{conv}(\lambda)$, given by the ratio $QE(\lambda)/A(\lambda)$ 
at near-normal incidence. As an example, in Fig. \ref{fig:QE(theta)_scint-air}
the predicted $QE(\theta)$ is plotted, for the ETL 9102B in scintillator
and air, at the wavelength of $410\, nm$. The curves are calculated
by rescaling the effective absorption in such a way that the QE in
air at normal incidence is 27 \%. It is observed that both curves have
a plateau, extending up to $\theta\simeq40^{\circ}$ in scintillator
and $\theta\simeq50^{\circ}$ in air. The QE in liquid is expected to be
slightly higher than in air, because of the Fresnel reflection at
the window. As in Fig. \ref{fig:RA_theta-lam}, PMTs in liquids
have a peak sensitivity in the range $45^{\circ}\lesssim\theta\lesssim75^{\circ}$.
On the other hand, PMTs in air lose sensitivity above $\sim50^{\circ}$,
because the Fresnel reflection from the glass window increases.

\subsection{\label{sub:QE_qualitative}Qualitative Understanding of the PMT Sensitivity}

In this section we show that the model is also able to explain qualitative
features of the wavelength dependence of the PMT QE. The latter is related 
to the photocathode absorption\footnote{For this qualitative discussion we use 
the simple absorption, instead of the effective absorption.}, 
as expressed in Eq. \ref{eq:QE}. 
Fig. \ref{fig:abs(lam)_air} shows the wavelength dependence of the absorption, 
deduced from our model (central solutions) for PMTs in air at near normal incidence.
The figure shows that the green-enhanced bialkali photocathode has
a plateau of maximum absorption extending further into the ``green''
visible region, compared to the blue-sensitive bialkali. Absorption is largely
determined by the imaginary part of the refractive index, so that
the features of the calculated spectra reflect the shape of the $k(\lambda)$
functions in Figs. \ref{fig:bluebialkali_N-K} and \ref{fig:greenbialkali_N-K}.
There it is also found that the imaginary part of the \emph{RbCsSb} refrective
index peaks at longer wavelengths compared to \emph{KCsSb}. 

We note that the correct prediction of the sensitivity of the two
photocathodes, though only qualitative, is non trivial, because it
is based exclusively on the study of the reflected light. However, 
the calculated absorption spectra are in disagreement with the reported 
QE (see Fig. \ref{fig:QE-spectra}), especially in the tail, where the QE drops
to zero, while the model predicts only a factor of $\sim3$ reduction.
This may be explained by postulating that the conversion factor $P_{conv}(\lambda)$
in Eq. \ref{eq:QE} vanishes at long wavelengths: photoelectrons
are indeed produced, but with insufficient kinetic energy to escape
the photocathode. Lay also comes to the same conclusion in \cite{Lay}.

\section{Conclusions}

We have investigated the optical properties in the
visible range of two test ETL PMTs, 9102B and 9902B, equipped with
a ``blue-sensitive'' \emph{KCsSb} and a ``green-enhanced'' \emph{RbCsSb}
bialkali photocathode, respectively. This study includes the spectroscopic
measurement of the absolute reflectance at near normal incidence, 
of the polarization-dependent relative reflectance at variable angles, 
and of the change in the light polarization upon reflection off the 
photocathode. 
We have fitted the experimental data with a model based on the optics 
of thin films to determine the photocathode thickness
and its wavelength-dependent complex refractive index. Degenerate
solutions appeared, however the fit converges to a unique solution once the 
complex refractive index of the photocathode is fixed at a single wavelength, 
for example to the value determined by other authors at $\lambda=442\, nm$. 
Based on this observation, we could predict the optical properties of both 
PMTs in the wavelength region $[380\, nm,\,680\, nm]$, and for any 
angle of incidence and coupling. As an example, we have studied 
in detail the case of a \emph{KCsSb} PMT coupled to a scintillator. 
The model predicts that the reflectance as a function of the angle 
of incidence on the PMT is nearly constant for 
$0^{\circ}<\theta\lesssim40^{\circ}$, with values of $R\sim15\,\%-25\,\%$, 
depending on the wavelength. It was shown that in this angular range 
the predictions of the model are very robust, since they are strongly 
constrained by the direct measurement of the PMT reflectance. 
Above the critical angle ($\theta\gtrsim42^{\circ}$)
the degeneracy is broken and the uncertainty of the model is larger.
However, also in this case quantitative predictions are made, at the
level of $\lesssim10\,\%$ precision ($2\sigma$). As another application
of this study, through the calculated angular dependence of the photocathode
absorption, the quantum efficiency of a PMT, normally measured
in air for near-normal illumination, can be predicted for any angle
and coupling. For the case of PMT in scintillator, we found that
the PMT has nearly constant sensitivity for $\theta<\theta_{c}$,
a $\sim20\,\%$ higher efficiency for $45^{\circ}\lesssim\theta\lesssim75^{\circ}$
and a vanishing efficiency for $\theta\rightarrow90^{\circ}$, where
the PMT reflectance goes to $100\,\%$. The sensitivity of a PMT in
air is expected to have a plateau for $\theta\lesssim50^{\circ}$
and then to decrease monotonically down to $0\,\%$ for $\theta\rightarrow90^{\circ}$,
due to the Fresnel reflection at the glass window.

The description of light interaction with the bialkali PMTs reported
in this paper covers most of the experimental conditions in which
these devices are used. Therefore this model is suitable for implementation
in photon-tracing Monte Carlo codes, through the formulas \ref{eq:R&T_amplitudes_thin_film}
- \ref{eq:Abs_thin_film}, the input optical parameters $n(\lambda)$,
$k(\lambda)$ shown in Figs. \ref{fig:bluebialkali_N-K}, \ref{fig:greenbialkali_N-K} and
listed in Appendix \ref{Appendix_A}, and the typical photocathode thickness reported in Tables 
\ref{tab:blue_fitpar} and \ref{tab:green_fitpar}.

\section*{Acknowledgments}

We would like to acknowledge the helpful support of Ron Stubberfield
from ETL for providing the test PMT samples and many useful documentation.
We thank Dr. M.E. Moorhead for helpful discussions. We are in debt
with Prof. Grunze of the Institute for applied physical chemistry
of the University of Heidelberg for the use of the ellipsometer, and
with Dr. Thimo Bastuck for the precious assistance in the ellipsometric
measurements. We thank Dr. F. Dalnoki-Veress for careful reading of the
manuscript and helpful suggestions.

\hfill
\newpage

\hfill
\newpage

\hfill
\newpage

\hfill
\newpage

\hfill
\newpage

\hfill
\newpage

\hfill
\newpage

\hfill
\newpage

\hfill
\newpage

\hfill
\newpage

\hfill
\newpage

\hfill
\newpage

\hfill
\newpage

\hfill
\clearpage
\newpage

\begin{table}
\caption{\textsl{\small \label{tab:blue_fitpar}Best fit photocathode
thickness, VASRA normalization factors (scale factors for the model
to fit data) and $\Delta$ phase shift for the ETL 9102B sample. }}
\begin{center}\begin{sideways}
\begin{tabular}{c|c|c|c|c|c|c|c|c}
&
d&
\multicolumn{6}{c|}{VASRA normalization factors}&
$\Delta$-shift\tabularnewline
&
(nm)&
\multicolumn{6}{c|}{}&
(deg)\tabularnewline
\hline
&
&
$p-45^{\circ}$ &
$p-55^{\circ}$ &
$p-65^{\circ}$ &
$s-45^{\circ}$ &
$s-55^{\circ}$ &
$s-65^{\circ}$ &
\tabularnewline
\hline 
central solution&
$20.0$&
$1.09$&
$1.00$&
$0.97$&
$1.18$&
$1.09$&
$1.04$&
$-9.5$\tabularnewline
\hline 
``upper'' solution&
$16.5$&
$1.05$&
$0.94$&
$0.90$&
$1.19$&
$1.10$&
$1.04$&
$-11.0$\tabularnewline
\hline 
``lower'' solution&
$25.0$&
$1.14$&
$1.09$&
$1.10$&
$1.17$&
$1.08$&
$1.03$&
$-7.9$\tabularnewline
\end{tabular}
\end{sideways}\end{center}

\end{table}
\hfill
\newpage

\begin{table}
\caption{\textsl{\small \label{tab:green_fitpar}Best fit photocathode
thickness, VASRA normalization factors (scale factors for the model
to fit data) and $\Delta$ phase shift for the ETL 9902B sample.}}

\begin{center}\begin{sideways}
\begin{tabular}{c|c|c|c|c|c|c|c|c}
&
d&
\multicolumn{6}{c|}{VASRA normalization factors}&
$\Delta$-shift\tabularnewline
&
(nm)&
\multicolumn{6}{c|}{}&
(deg)\tabularnewline
\hline
&
&
$p-45^{\circ}$ &
$p-55^{\circ}$ &
$p-65^{\circ}$ &
$s-45^{\circ}$ &
$s-55^{\circ}$ &
$s-65^{\circ}$ &
\tabularnewline
\hline 
central solution&
$23.4$&
$1.17$&
$1.07$&
$0.98$&
$1.40$&
$1.25$&
$1.13$&
$-11.2$\tabularnewline
\hline 
``upper'' solution&
$19.4$&
$1.14$&
$1.03$&
$0.93$&
$1.41$&
$1.25$&
$1.13$&
$-12.3$\tabularnewline
\hline 
``lower'' solution&
$29.0$&
$1.21$&
$1.14$&
$1.07$&
$1.40$&
$1.24$&
$1.12$&
$-9.9$\tabularnewline
\end{tabular}
\end{sideways}\end{center}

\end{table}

\hfill
\clearpage
\newpage

\appendix

\section{\label{Appendix_A}Best Fit Optical Parameters}

We report in the following table the best fit refractive indices for
the two photocathodes, from $380\, nm$ to $680\, nm$ in $15\, nm$
steps, as implemented in the minimization code. The central solutions
are shown, which correspond to a fixed $n=2.70$ and $k=1.50$ for
the \emph{KCsSb} photocathode, and $n=2.50$ and $k=1.35$ for \emph{RbCsSb}.
The relevant best fit thicknesses are $20.0\, nm$ and $23.4\, nm$, for the 
\emph{KCsSb} and \emph{RbCsSb} photocathode, respectively.

\begin{center}\begin{longtable}{c||c|c||c|c||}
&
\multicolumn{2}{|c||}{KCsSb}&
\multicolumn{2}{c||}{RbCsSb}\tabularnewline
\hline
\hline 
$\lambda\,(nm)$&
n&
k&
n&
k\tabularnewline
\hline
\endhead
\hline 
380&
1.92&
1.69&
2.07&
1.22\tabularnewline
\hline 
395&
2.18&
1.69&
2.22&
1.17\tabularnewline
\hline 
410&
2.38&
1.71&
2.30&
1.21\tabularnewline
\hline 
425&
2.61&
1.53&
2.40&
1.27\tabularnewline
\hline 
440&
2.70&
1.50&
2.50&
1.35\tabularnewline
\hline 
455&
2.87&
1.44&
2.63&
1.40\tabularnewline
\hline 
470&
3.00&
1.34&
2.81&
1.41\tabularnewline
\hline 
485&
3.00&
1.11&
2.99&
1.37\tabularnewline
\hline 
500&
3.00&
1.06&
3.16&
1.21\tabularnewline
\newpage
\hline 
515&
3.09&
1.05&
3.22&
1.04\tabularnewline
\hline 
530&
3.26&
0.86&
3.21&
0.90\tabularnewline
\hline 
545&
3.20&
0.63&
3.23&
0.86\tabularnewline
\hline 
560&
3.12&
0.53&
3.32&
0.76\tabularnewline
\hline 
575&
3.06&
0.46&
3.39&
0.59\tabularnewline
\hline 
590&
3.01&
0.42&
3.28&
0.46\tabularnewline
\hline 
605&
2.98&
0.38&
3.22&
0.43\tabularnewline
\hline 
620&
2.96&
0.37&
3.18&
0.40\tabularnewline
\hline 
635&
2.95&
0.35&
3.15&
0.38\tabularnewline
\hline 
650&
2.95&
0.34&
3.14&
0.37\tabularnewline
\hline 
665&
2.95&
0.34&
3.14&
0.37\tabularnewline
\hline 
680&
2.96&
0.33&
3.13&
0.35\tabularnewline
\end{longtable}\end{center}

\end{document}